\documentclass{ws-procs9x6-cpt16}
\begin{document}

\newcommand{\refeq}[1]{(\ref{#1})}
\def\etal {{\it et al.}}
%any other macros go here 

\def\al{\alpha}
\def\be{\beta}
\def\ga{\gamma}
\def\de{\delta}
\def\ep{\epsilon}
\def\ve{\varepsilon}
\def\ze{\zeta}
\def\et{\eta}
\def\th{\theta}
\def\vt{\vartheta}
\def\io{\iota}
\def\ka{\kappa}
\def\la{\lambda}
\def\vpi{\varpi}
\def\rh{\rho}
\def\vr{\varrho}
\def\si{\sigma}
\def\vs{\varsigma}
\def\ta{\tau}
\def\up{\upsilon}
\def\ph{\phi}
\def\vp{\varphi}
\def\ch{\chi}
\def\ps{\psi}
\def\om{\omega}
\def\Ga{\Gamma}
\def\De{\Delta}
\def\Th{\Theta}
\def\La{\Lambda}
\def\Si{\Sigma}
\def\Up{\Upsilon}
\def\Ph{\Phi}
\def\Ps{\Psi}
\def\Om{\Omega}
\def\cA{{\cal A}}
\def\cB{{\cal B}}
\def\cC{{\cal C}}
\def\cE{{\cal E}}
\def\cl{{\mathcal L}}
\def\cL{{\mathcal L}}
\def\cO{{\cal O}}
\def\cP{{\cal P}}
\def\cR{{\cal R}}
\def\cV{{\cal V}}
\def\mn{{\mu\nu}}

\def\fr#1#2{{{#1} \over {#2}}}
\def\half{{\textstyle{1\over 2}}}
\def\quar{{\textstyle{1\over 4}}}
\def\frac#1#2{{\textstyle{{#1}\over {#2}}}}

\def\vev#1{\langle {#1}\rangle}
\def\bra#1{\langle{#1}|}
\def\ket#1{|{#1}\rangle}
\def\bracket#1#2{\langle{#1}|{#2}\rangle}
\def\expect#1{\langle{#1}\rangle}
\def\norm#1{\left\|{#1}\right\|}
\def\abs#1{\left|{#1}\right|}

\def\lsim{\mathrel{\rlap{\lower4pt\hbox{\hskip1pt$\sim$}}
   \raise1pt\hbox{$<$}}}
\def\gsim{\mathrel{\rlap{\lower4pt\hbox{\hskip1pt$\sim$}}
   \raise1pt\hbox{$>$}}}
\def\sqr#1#2{{\vcenter{\vbox{\hrule height.#2pt
        \hbox{\vrule width.#2pt height#1pt \kern#1pt
        \vrule width.#2pt}
        \hrule height.#2pt}}}}
\def\square{\mathchoice\sqr66\sqr66\sqr{2.1}3\sqr{1.5}3}

\def\prt{\partial}

\def\etal{{\it et al.}}

\def\pt#1{\phantom{#1}}
\def\ni{\noindent}
\def\ol#1{\overline{#1}}

\def\nsc#1#2#3{\om_{#1}^{{\pt{#1}}#2#3}}
\def\lsc#1#2#3{\om_{#1#2#3}}
\def\usc#1#2#3{\om^{#1#2#3}}
\def\lulsc#1#2#3{\om_{#1\pt{#2}#3}^{{\pt{#1}}#2}}

\def\tor#1#2#3{T^{#1}_{{\pt{#1}}#2#3}}

\def\vb#1#2{e_{#1}^{{\pt{#1}}#2}}
\def\ivb#1#2{e^{#1}_{{\pt{#1}}#2}}
\def\uvb#1#2{e^{#1#2}}
\def\lvb#1#2{e_{#1#2}}

\newcommand{\beq}{\begin{equation}}
\newcommand{\eeq}{\end{equation}}
\newcommand{\bea}{\begin{eqnarray}}
\newcommand{\eea}{\end{eqnarray}}
\newcommand{\bit}{\begin{itemize}}
\newcommand{\eit}{\end{itemize}}
\newcommand{\rf}[1]{(\ref{#1})}

\title{Background Fields and Gravity}

\author{R.\ BLUHM}

\address{Department of Physics, Colby College,\\
Waterville, ME 04901, USA}

\begin{abstract}
Gravitational theories with fixed background fields
break diffeomorphism invariance.
This breaking can be spontaneous or explicit.
A brief summary of the main consequences of
these types of breaking is presented.
\end{abstract}

\bodymatter

\section{Introduction}	

Violation of local Lorentz invariance is a feature of many
theoretical models that attempt to merge General Relativity (GR)
with quantum physics and the Standard Model (SM) of particle physics.
The Standard-Model Extension
(SME) is the theoretical framework used by theorists
and experimentalists to search for possible signals of 
Lorentz violation.\cite{sme,smereview}
The SME is constructed as the general observer-independent
effective field theory that incorporates breaking of local Lorentz invariance.
The Lorentz-violating interactions in the SME Lagrangian consist 
of contractions of gravitational and SM fields with fixed 
background fields referred to as SME coefficients.
Experiments with sensitivity to Lorentz breaking make measurements
that place bounds on the SME coefficients.\cite{data}

In the gravity sector, the SME assumes general coordinate invariance
so as to ensure observer and coordinate independence.
At the same time, the SME coefficients act as fixed background 
fields that break both local Lorentz invariance (in local frames) 
as well as spacetime diffeomorphism invariance (in spacetime frames).
These breakings can occur either through spontaneous symmetry breaking, 
where the backgrounds form as vacuum values of dynamical fields,
or through explicit breaking.
With explicit breaking,
the background fields appear directly in the Lagrangian as
pre-existing nondynamical fields.
The presence of such nondynamical backgrounds can lead to 
conflicts with the Bianchi identities,
while spontaneous breaking evades this problem.\cite{ak}
To avoid the conflicts associated with explicit breaking, 
the SME assumes that the background
fields arise from a process of spontaneous symmetry breaking.

In the following section, a brief summary is given of what the consequences
are when a gravitational theory includes a fixed background field.\cite{rb1}
In particular, the differences between spontaneous and explicit 
symmetry breaking are considered in relation to the Bianchi identities.
The counting of degrees of freedom is considered as well,
and comparisons are made to the symmetry-preserving case of 
Einstein's GR.

\section{Spontaneous versus Explicit Breaking}	

For simplicity in this analysis, 
Riemann spacetime described by an Einstein-Hilbert term can be used, 
and the Lorentz-breaking sector can be restricted to a potential term
involving only a generic background $\bar k_{\la\mu\nu\cdots}$ and the metric.
The action is then given as
\beq
S = \int \sqrt{-g} \, d^4 x \left( \fr 1 {16 \pi G} R - {\cal U}(g_{\mu\nu}, \bar k_{\la\mu\nu\cdots}) \right) .
\label{action}
\eeq
The fixed background breaks diffeomorphism invariance.
Local Lorentz symmetry is broken as well;
however, its treatment typically involves using a vierbein formalism.
By restricting to a metric description it suffices in this context to focus only on
the diffeomorphism breaking.

A potential conflict involving the Bianchi identities can be seen
by performing a general coordinate transformation, $x^\mu \rightarrow x^\mu + \xi^\mu$,
and using the fact that the action is invariant.
This gives the off-shell condition:
\beq
\int d^4x \sqrt{-g} \left[ (D_\mu T^{\mu\nu})\xi_\nu 
+ \fr {\de {\cal U}} {\de \bar k_{\la\mu\nu \cdots}} {\cal L}_\xi \bar k_{\la\mu\nu \cdots} \right] = 0 .
\label{GCT}
\eeq
In deriving this condition,
the vectors $\xi^\mu$ are assumed to vanish on the boundary,
and the contracted Bianchi identities, $D_\mu G^{\mu\nu} = 0$, are used.
When the Einstein equations hold, $D_\mu T^{\mu\nu}$
must vanish, which is also a result of the contracted Bianchi identities.
Consistency of the theory requires that the integral of the 
second term in \rf{GCT} must therefore vanish on shell.

With explicit diffeomorphism breaking, the background $\bar k_{\la\mu\nu\cdots}$
is nondynamical, and the variation of the potential ${\cal U}$ with respect
to it need not vanish.
In addition, the Lie derivative, ${\cal L}_\xi \bar k_{\la\mu\nu\cdots}$, is assumed to be nonzero,
since the background breaks diffeomorphism invariance.
As a result, the integrand in \rf{GCT} is nonzero, 
and there is potentially a conflict with the Bianchi identities if the integral does not vanish.

In contrast, when the breaking is spontaneous,
the background arises as a dynamical vacuum solution given
as a vacuum expectation value, $\bar k_{\la\mu\nu\cdots} = \vev{k_{\la\mu\nu\cdots}}$.
In this case, the variation of the potential ${\cal U}$ with respect to the
background vanishes, and there is no conflict with the Bianchi identities.

It is primarily for this reason, and to avoid having pre-existing nondynamical fields
in a gravitational theory,
that the SME assumes the symmetry breaking is spontaneous.

However, the case of explicit breaking can be examined further, 
and it is found that evasion of the conflict with the Bianchi identities can occur.
This is because the integrand in \rf{GCT} can be shown to equal a
total divergence:
\beq
(D_\mu T^{\mu\nu}) \xi_\nu + 
 \fr {\de {\cal U}} {\de \bar k_{\la\mu\nu\cdots}}({\cal L}_{\xi} \bar k_{\la\mu\nu\cdots} )
= D_\mu \left(  2 \fr {\de {\cal U}} {\de g^{\al\be}} g^{\mu\al} \xi^\be \right) .
\label{deriv}
\eeq
Therefore, as long as the right-hand side of this expression exists and does not vanish, 
then the integral in \rf{GCT} equals zero
and there is no conflict with the Bianchi identities.

Examples of theories where the total derivative does not exist are known,
and such theories are either inconsistent or only exist for 
particular choices of spacetime geometry or when additional conditions apply.\cite{rb1,rb2}
On the other hand, there are large classes of theories where
the total derivative does exist,
and for these theories there is no conflict with the Bianchi identity.

In the absence of dynamics for the background field,
the reason why $D_\mu T^{\mu\nu} = 0$ can hold is
because the metric has four additional degrees of freedom
when diffeomorphism invariance is explicitly broken.
As long as the metric has sufficient coupling with
the nondynamical background $\bar k_{\la\mu\nu\cdots}$,
then the four extra modes in the metric can conspire together
to impose the four equations $D_\mu T^{\mu\nu} = 0$.
This behavior does not occur in GR or when diffeomorphism
invariance is spontaneously broken.
In those cases, covariant energy-momentum conservation
holds as a result of the equations of motion for the
dynamical fields,
and four degrees of freedom in the metric remain gauge degrees of freedom.

\section{Summary and Conclusions}

Gravitational theories with background fields break diffeomorphism invariance.
If the breaking is explicit, it is due to the appearance of pre-existing
nondynamical background fields in the action.
However, such a theory must still be invariant under general coordinate
transformations to maintain observer independence.
As a result, the condition, $D_\mu T^{\mu\nu} = 0$, which follows
from Einstein's equations and the contracted Bianchi identity, can 
then hold only if the integral of the right-hand term in \rf{GCT} equals zero.
From this condition different outcomes can occur,
depending in general on the extent to which the metric
couples with the background.
In cases where there is little or no coupling, 
the theory is either inconsistent
or the spacetime geometry must be restricted.
However, in cases where the metric couples in a way that allows the
total-derivative on the right-hand side of \rf{deriv} to exist,
then the conflict with the Bianchi identities is avoided.
The theory then has four extra degrees of freedom in the metric,
and it is these degrees of freedom that impose the 
conditions $D_\mu T^{\mu\nu} = 0$.

However, when the symmetry breaking is spontaneous,
the background $\bar k_{\la\mu\nu\cdots}$ arises dynamically.
The variations of the potential ${\cal U}$ with respect to the background
then vanish in \rf{GCT}, and there is no conflict with the Bianchi identities.
Theories with spontaneous diffeomorphism breaking therefore maintain many of
the usual features that occur in GR.
For example, when the Nambu-Goldstone and massive excitations for the background
field are included, diffeomorphism invariance of the action is restored,\cite{rbak}
and the metric again has four gauge degrees of freedom.
In this case, the equations $D_\mu T^{\mu\nu} = 0$ follow as a
result of the dynamics of the background field.

The SME assumes that the background coefficients arise at a
fundamental level from spontaneous breaking of local Lorentz and
diffeomorphism invariance. 
They are therefore not pre-existing and instead arise as vacuum values of dynamical fields.
In developing weak limits and a Post-Newtonian framework from the gravity
sector of the SME,
the additional Nambu-Goldstone and massive excitations for the background fields
must be taken into account.
Remarkably, in many cases of interest,
the conditions of diffeomorphsim invariance and the Bianchi identities
allow the background excitations to be
eliminated in terms of expressions involving only the metric, curvature,
and their derivatives.\cite{qbak06,akjt}
It is in this way that a useful phenomenological framework for testing Lorentz
violation in gravity is derived.


\begin{thebibliography}{xx}


\bibitem{sme}
D.\ Colladay and V.A.\ Kosteleck\'y,
Phys.\ Rev.\ D {\bf 55}, 6760 (1997);
Phys.\ Rev.\ D {\bf 58}, 116002 (1998).

\bibitem{smereview}
For reviews of the SME, see
R.\ Bluhm, hep-ph/0506054;
arXiv:1302.1150.

\bibitem{data}
V.A.\ Kosteleck\'y and N.\ Russell,
arXiv:0801.0287v9.

\bibitem{ak}
V.A.\ Kosteleck\'y,
Phys.\ Rev.\ D {\bf 69}, 105009 (2004).

\bibitem{rb1}
R.\ Bluhm,
Phys.\ Rev.\ D {\bf 91}, 065034 (2015).

\bibitem{rb2}
R.\ Bluhm,
Phys.\ Rev.\ D {\bf 92}, 085015 (2015).

\bibitem{rbak}
R.\ Bluhm and V.A.\ Kosteleck\'y,
Phys.\ Rev.\ D {\bf 71}, 065008 (2005);
R.\ Bluhm, S.-H.\ Fung, and V.A.\ Kosteleck\'y,
Phys.\ Rev.\ D {\bf 77}, 065020 (2008).

\bibitem{qbak06}
Q.G. Bailey and V.A.\ Kosteleck\'y,
Phys.\ Rev.\ D {\bf 74}, 045001 (2006).

\bibitem{akjt}
V.A.\ Kosteleck\'y and J.D.\ Tasson,
Phys.\ Rev.\ Lett.\ {\bf 102}, 010402 (2009);
Phys.\ Rev.\ D {\bf 83}, 016013 (2011).


\end{thebibliography}
\end{document}